\documentclass[]{aastex631}

\usepackage{array}
\usepackage{multirow}

\begin{document}

\title{SZ Lyncis: A Non-Accreting Neutron Star–$\delta$ Scuti Binary Candidate Discovered via Dynamics and Asteroseismology}

\email{liping@ynao.ac.cn}
\author{Ping Li}
\affiliation{Yunnan Observatories, Chinese Academy of Sciences (CAS), 650216 Kunming, P.R. China}
\affiliation{University of Chinese Academy of Sciences, No.1 Yanqihu East Rd, Huairou District, Beijing, P.R.China 101408}	

\author{Li-Ying Zhu}
\affiliation{Yunnan Observatories, Chinese Academy of Sciences (CAS), 650216 Kunming, P.R. China}
\affiliation{University of Chinese Academy of Sciences, No.1 Yanqihu East Rd, Huairou District, Beijing, P.R.China 101408}	

\author{Wen-Ping Liao}
\affiliation{Yunnan Observatories, Chinese Academy of Sciences (CAS), 650216 Kunming, P.R. China}
\affiliation{University of Chinese Academy of Sciences, No.1 Yanqihu East Rd, Huairou District, Beijing, P.R.China 101408}



\begin{abstract}
Neutron stars (NSs) are traditionally discovered through radio, X-ray, or gamma-ray observations, but optical time-domain surveys can unveil non-accreting NSs in wide binaries. Here we report a NS candidate in the single-lined binary SZ~Lyncis, identified through a combination of asteroseismology, spectroscopy, pulsation timing, and astrometry. The visible $\delta$ Scuti primary has a mass of $M_1 = 1.83_{-0.01}^{+0.06}~\mathrm{M_{\odot}}$ from asteroseismic modeling. With the orbital inclination ($i = 38.67 \pm 0.29^\circ$) from the astrometric data of Gaia and Hipparcos, we obtain companion masses of $M_2 = 1.76_{-0.042}^{+0.042}~\mathrm{M_{\odot}}$ (radial velocity) and $M_2 = 2.07_{-0.045}^{+0.045}~\mathrm{M_{\odot}}$ (timing variations). The companion's mass exceeds the Chandrasekhar limit and lies in the NS range. Multiple arguments rule out alternatives: the astrometric mass function and the spectral energy distribution, which shows no extra light, together exclude any luminous companion; the mass and lack of Balmer absorption rule out white dwarfs (WDs); the system's age ($1.25$~Gyr) disfavors a double WD; and the mass is too low for a black hole. The wide, low-eccentricity orbit and absence of accretion signatures are consistent with a quiescent NS. SZ~Lyn has the potential to be the first $\delta$ Scuti binary with a NS candidate identified through asteroseismology and dynamics, demonstrating the potential of this approach to uncover non-accreting compact objects.
\end{abstract}



\section{Introduction} \label{sec:intro}
The mass distribution and physical properties of neutron stars (NSs) encode crucial information about stellar evolution history and Galactic chemical enrichment \citep{2012ApJ...757...55O,2022NatAs...6.1203Y}. While typically detected through electromagnetic emission – including rapidly rotating, highly magnetized radio pulsars \citep{2008LRR....11....8L}, accreting X-ray binaries, gamma-ray pulsars \citep{2010ApJS..187..460A}, and isolated thermally emitting NSs \citep{2007Ap&SS.308..181H} – with merging NS systems additionally detectable via LIGO/Virgo gravitational-wave observatories \citep{2017PhRvL.119p1101A}, these methods possess inherent limitations. Radio, X-ray, and gamma-ray detection requires either favorable beaming geometry or ongoing accretion processes \citep{2022NatAs...6.1203Y}, leaving non-accreting and radio-quiet NS populations largely undetected \citep{1996A&ARv...7..209C}.

$\delta$ Scuti variable stars (A--F spectral types) occupy the main-sequence region intersecting the classical Cepheid instability strip in the Hertzsprung-Russell diagram. Their oscillations, driven by the $\kappa$ mechanism in the He\,{\sc ii} partial ionization zone, produce stable periodic signals with periods of 15 minutes to 8 hours \citep{2010aste.book.....A, 1993A&A...271..482B}. High-amplitude $\delta$ Scuti stars (HADS)—defined by photometric variations $\geq$ 0.1 mag and typically slow rotation ($v \sin i \leq 30$ km s$^{-1}$)—function as exceptional celestial chronometers \citep{2026arXiv260112999N}. Their remarkable pulsational stability enables precise detection of minuscule period variations induced by orbital motion through the light-travel time effect (LTTE) \citep{1952ApJ...116..211I}.

SZ~Lyncis (SZ~Lyn; TIC~192939152) is a well-studied HADS in a single-line spectroscopic binary system, with a mean magnitude of $\langle V \rangle = 9.1$~mag where only the brighter component (SZ~Lyn~A) is spectroscopically observable while the fainter companion (SZ~Lyn~B) remains undetected \citep{2004CoAst.144...26G,1981A&A....98..198B}. The fundamental pulsation period was first established by \cite{1968AJ.....73...29B} and subsequently refined by \cite{2004CoAst.144...26G}, with orbital studies showing that the linear ephemeris for pulsation maxima is modulated by long-term orbital motion through the LTTE \citep{1975AJ.....80...48B}. Both pulsation and orbital parameters have been progressively improved through later analyses \citep{1988AJ.....95.1534M,1988Ap&SS.149...73P,2013PASJ...65..116L,2004CoAst.144...26G}. Recently, \cite{2021MNRAS.502..541A} identified in SZ~Lyn~A a dominant radial mode ($f_1=8.296$~d$^{-1}$), two non-radial p-modes ($f_2=14.535$~d$^{-1}$, $f_3=32.620$~d$^{-1}$), and a potential g-mode ($f_4=4.584$~d$^{-1}$). The physical parameters of SZ~Lyn~A determined by previous authors and our work are summarized in Table~\ref{Tab:phy_para}, though the mass and radius of SZ~Lyn~A have not been accurately determined in earlier studies.
\begin{table}[ht]
\centering
\caption{Physical parameters of SZ Lyn A.}\label{Tab:phy_para}
\begin{tabular}{@{}lcc@{}}
\hline
\noalign{\smallskip}
Parameter & Value & Reference \\
\noalign{\smallskip}\hline
\noalign{\smallskip}
\multirow{4}{*}{$T_{\mathrm{eff}}$ (K)}
        & 7173 & LAMOST (this paper) \\
        & 7799 & \textit{Gaia} \\
        & 6750 & \cite{2024BSRSL..93..470A} \\ 
\noalign{\smallskip}
\multirow{2}{*}{$\log(g)$}
        & 3.50 & \cite{2024BSRSL..93..470A} \\
        & 3.87 & LAMOST (this paper)\\
\noalign{\smallskip}
\multirow{2}{*}{Mass (M$_\odot$)}
        & $1.57^{+0.17}_{-0.66}$ & \cite{1984MNRAS.208..853F} \\
        & 1.7-2.0 & \cite{2021MNRAS.502..541A} \\
\noalign{\smallskip}

\noalign{\smallskip}
\multirow{2}{*}{Radius (R$_\odot$)}
        & $2.76^{+0.11}_{-0.46}$ & \cite{1984MNRAS.208..853F} \\
        & 2.80 & Bardin \& Imbert (1981) \\
\noalign{\smallskip}
Parallax (mas) & $2.49 \pm 0.07$ & \textit{Gaia} \\
\noalign{\smallskip}\hline
\end{tabular}
\label{tab:lit_params}
\end{table}

Two key developments now enable a definitive reassessment of this system. First, high-precision, continuous photometry from the Transiting Exoplanet Survey Satellite (TESS) \citep{2015JATIS...1a4003R} provides exquisite characterization of the primary star's pulsation modes, yielding significantly improved determinations of mass and radius for SZ~Lyn~A along with precise times of maximum light. Second, by incorporating the orbital inclination derived from Hipparcos and Gaia data, we can estimate the mass of SZ~Lyn~B through dynamical analysis combining radial velocity (RV) measurements and the LTTE, and determine its nature.

Building on these advances, our comprehensive analysis combines TESS photometry with times of maximum light and RV measurements to reveal an astrophysical treasure: a wide, low-eccentricity orbit harboring a compact companion. Mass estimates of $M_2 = 1.76_{-0.042}^{+0.042}$ M$_{\odot}$ (RV method) and $M_2 = 2.07_{-0.045}^{+0.045}$ M$_{\odot}$ (LTTE method) indicate that SZ Lyn B is a NS. If this is correct, it would be the first NS known to orbit a $\delta$ Scuti star at a distance of 397 pc.

The system exhibits notable characteristics: it ranks among the rare binaries hosting both a main-sequence pulsator and a NS candidate, enabling potential asteroseismology in strong gravity; its wide, low-eccentricity orbit offers constraints for supernova kick models; and it pioneers the synergy of asteroseismology with dynamical methods (radial velocity and LTTE) for compact companion detection. Collectively, these features indicate SZ~Lyn may represent a scientifically valuable system, with potential implications for understanding stellar evolution and advancing compact star searches.

\section{Results and discussion} \label{sec:result}
\subsection{Asteroseismic properties of SZ~lyn~A}
Analysis of the TESS photometry confirms SZ~Lyn~A as a multi-periodic pulsator, revealing six significant frequencies including four previously reported by \cite{2021MNRAS.502..541A}. To provide complementary constraints on the system properties, we conducted asteroseismic modeling of the primary star. Using Modules for Experiments in Stellar Astrophysics (MESA) \citep{2011ApJS..192....3P,2013ApJS..208....4P,2015ApJS..220...15P,2018ApJS..234...34P}, we computed a grid of stellar evolutionary models, then derived their adiabatic pulsation frequencies with \textit{pulse\_adipls} \citep{2008Ap&SS.316..113C}. The optimal model, selected through frequency matching of all six observed pulsation modes following the identification framework of \cite{2021MNRAS.502..541A}, is presented in Table \ref{tab:model_para} (In Appendix).

The best-fit model (Table \ref{tab: para} in Appendix) places SZ~Lyn~A on the post-main sequence, yielding the following parameters: mass $M = 1.83^{+0.06}_{-0.01}$~M$_{\odot}$, effective temperature $T_{\rm eff} = 6791^{+51}_{-58}$~K, radius $R = 2.899^{+0.027}_{-0.000}$~R$_{\odot}$, luminosity $L = 16.111^{+0.721}_{-0.354}$~L$_{\odot}$, and age $=1.254^{+0.079}_{-0.024}$~Gyr. Notably, the derived mass and radius are consistent with the values in Table~\ref{Tab:phy_para} but exhibit significantly reduced uncertainties.

\subsection{The orbital ephemeris, radial velocity curve, and the mass function}
The orbital ephemeris of SZ~Lyn~A is:
\begin{equation}\label{equation 1}
T (\phi=0)= 2445156.600(HJD)+1181^d.5(1.4)\times N
\end{equation}
where the first term is the primary eclipse time (denoted as $T_0$), and the second term is the orbital period in day times the umber of orbital cycles $N$ \citep{1984A&AS...57..249B}, HJD is the Heliocentric Julian Date.

\begin{figure}[ht!]
	\plotone{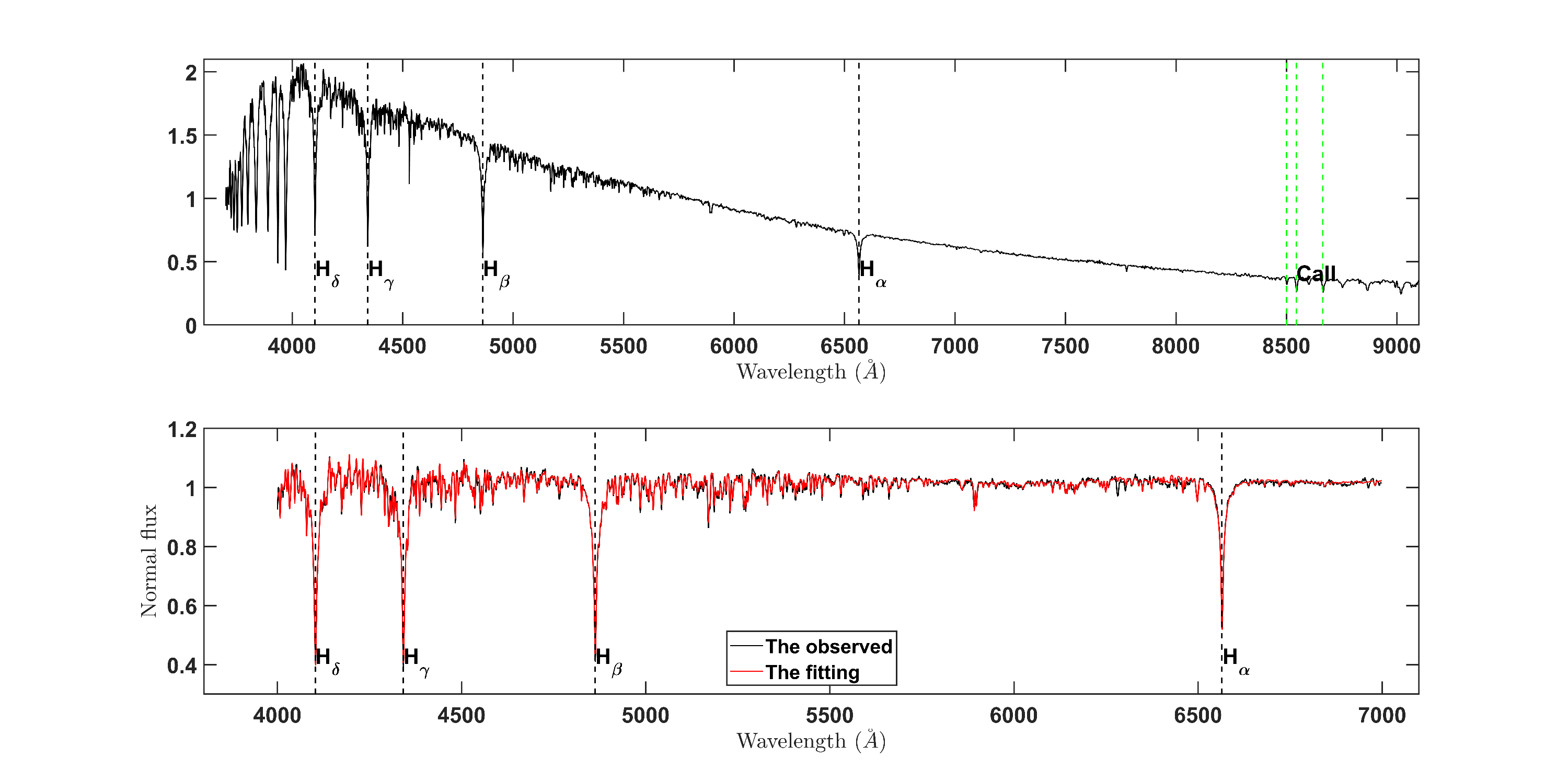}
	\caption{ The low-resolution spectrum of SZ Lyn observed with LAMOST. Upper panel: The original spectrum, the four Balmer absorbed lines H$_{\alpha}$, H$_{\beta}$, H$_{\gamma}$, and H$_{\delta}$ and the metal lines CaII are marked. Lower panel: The fitting model of A-type star for the spectrum of SZ Lyn.} \label{fig:spec}
\end{figure}

As shown in Figure \ref{fig:spec}, a low-resolution spectrum of SZ Lyn was taken from DR7 of LAMOST low-resolution single-epoch spectra. the University of Lyon Spectroscopic Analysis Software (ULySS) \citep{ 2009A&A...501.1269K} was used to obtain stellar atmospheric parameters: $T_{\rm teff}$=7173 ($\pm 146$) K, $\log g$=3.87 ($\pm 0.01$), [$Fe/H$]=-0.08 ($\pm 0.01$). The cross-corresponding function (CCF) method is used to obtained a RV value (32$\pm4$ km/s). We also collected 21 Rv values from  \cite{1984A&AS...57..249B}, covering one orbital period of SZ Lyn. According to Equation (1), The RV data are phase-folded.

The following equations\citep{2015PASP..127..567I} are used to fit the RV values
\begin{equation}\label{equation 2}
V_r=\gamma+K[\cos(\theta+\omega)+e \cos(\omega)],
\end{equation}
\begin{equation}\label{equation 3}
\theta(t)=2\arctan[\sqrt{\frac{1+e}{1-e}}\tan(\frac{E}{2})],
\end{equation}
\begin{equation}\label{equation 4}
E-e\sin(E)=\frac{2\pi}{P}(t-T_p),
\end{equation}
where $\gamma$, $K$, $\omega$, $e$, $E$, $P$, and $T_p$ are the center of mass velocity of binary system, the semi-amplitude of RV, the eccentricity of the orbit, the eccentric anomaly, the orbital period of binary system. and the time of periastron passage, respectively.

\begin{figure}[ht!]
	\plotone{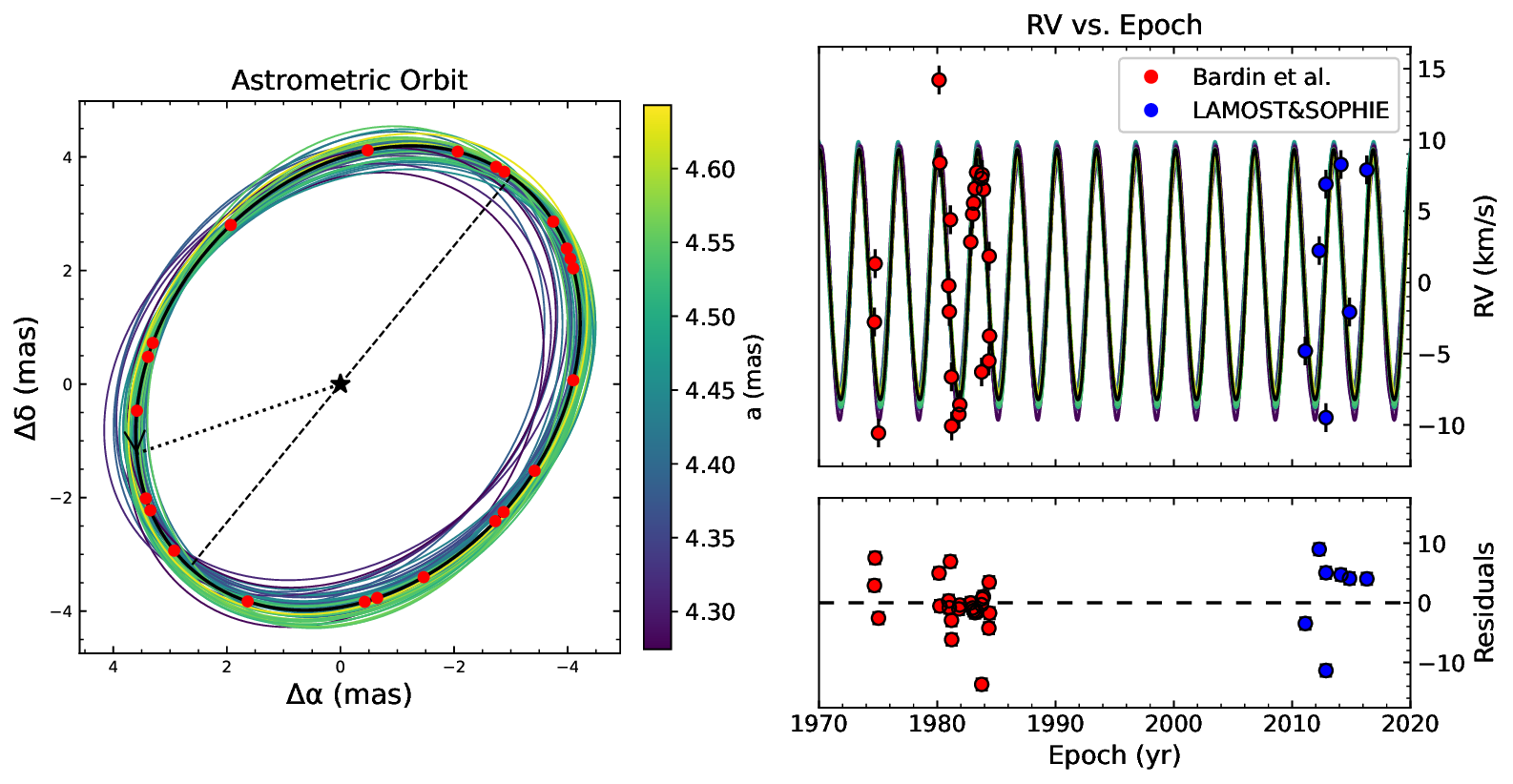}
	\caption{ Left panel: predicted observation times of SZ Lyn from the Gaia Observation Scheduling Tool (GOST) and Hipparcos HIP1 and 2. The black line shows the best-fitting orbit using the orvara code from our combined fit to the radial velocity data, Hipparcos astrometry, and the GOST scan time predictions. The red points show the model-predicted photocentre positions at the times when GOST predicts Gaia would have observed the source, computed from the best-fitting orbit. We do not have access to the actual measured Gaia astrometric positions (RA, Dec); only to the predicted scan times. The expected per-epoch astrometric uncertainties are approximately 0.2 mas in the along-scan direction. The predicted scan times demonstrate that the orbit has been sampled adequately. Right panel: the radial velocity curve of SZ Lyn.} \label{fig:rv}
\end{figure}

Figure \ref{fig:rv} shows that the fitting results of RV in $\gamma$=34.18 ($\pm$0.05) (km/s), $K$=9.51 ($\pm$0.09) (km/s), $\omega$=101.2 ($\pm$1.4)$^{\circ}$, $e$=0.186 ($\pm$0.009), $P$=1188.51 ($\pm$7.59) days and $T_p$=2452751.81 ($\pm$19.13), respectively.  Hence the mass function (lower mass limit) for SZ Lyn B is:
\begin{equation}\label{equation 5}
f(M_2)=\frac{(M_2\sin{i})^3}{(M_1+M_2)^2}=\frac{PK^3}{2\pi G}(1-e^2)^{3/2}=0.1031\pm0.0030 \rm M_{\odot}
\end{equation}
where $M_1$ and $M_2$ are the masses of SZ Lyn A and B, respectively, $i$ is the orbital inclination, and $G$ is the gravitational constant.

\subsection{The O-C analysis (LTTE) of the pulsation maximum}
The orbital ephemeris of SZ Lyn is:
\begin{equation}\label{equation 6}
T' = 2438124.39955(HJD)+0^d.12053491\times E
\end{equation}
where the first term is the pulsation maximum time (denoted as $T'_0$), and the second term is the pulsation period in day times the umber of pulsation cycles $E$ \citep{1988Ap&SS.149...73P}.

We obtained 413 pulsation maximum from the light curves observed by TESS. There are 202 pulsation maximum are collected from \cite{2022PhDT.........2A}, observed by $AAVSO$, $WASP$ ( $Wide$ $Angle$ $Search$ $for$ $Planets$) and $Abu$ \citep{2021MNRAS.502..541A}. In addition, we also collected 337 pulsation maximum from \cite{2003IBVS.5485....1A,2003A&A...402..733D,2004CoAst.144...26G,2003IBVS.5485....1A,2006IBVS.5701....1K,2010JAVSO..38...12S,2011JAVSO..39...23S,2012JAVSO..40..923S,2013JAVSO..41...85S,2007IBVS.5761....1H,2006IBVS.5731....1H,2009IBVS.5889....1H,2012IBVS.6026....1H,2010IBVS.5918....1H,2009IBVS.5874....1H,2005IBVS.5657....1H,2013IBVS.6048....1H,2013IBVS.6049....1W}. According to Equation \ref{equation 6}, pulsation maximum data are converted to $O-C$ and pulsation number cycles. The following equations\citep{1952ApJ...116..211I} are used to fit the $O-C$ values
\begin{equation}\label{equation 8}
 O-C= \Delta T_0 + \Delta P_0 \cdot E + A\left[ \sqrt{1 - e^2}\sin E^*\cos\omega + \cos E^*\sin\omega \right],
\end{equation}
and the Kepler’sequation:
\begin{equation}\label{equation 9}
M = E^* - e\sin E^* = \frac{2\pi(t - T)}{P}
\end{equation}
where $\Delta T_0$ and $\Delta P_0$ denote the revised epoch and period, $A$ is the $O\!-\!C$ semi-amplitude, $e$ the eccentricity, $\nu$ the true anomaly, $\omega$ the longitude of periastron, $M$ the mean anomaly, $T$ the periastron passage time, $t$ the observed pulsating maximum times, $P$ the orbital period, and $E^*$ the eccentric anomaly.
\begin{figure}[ht!]
    \centering
	\includegraphics[width=1.1\columnwidth]{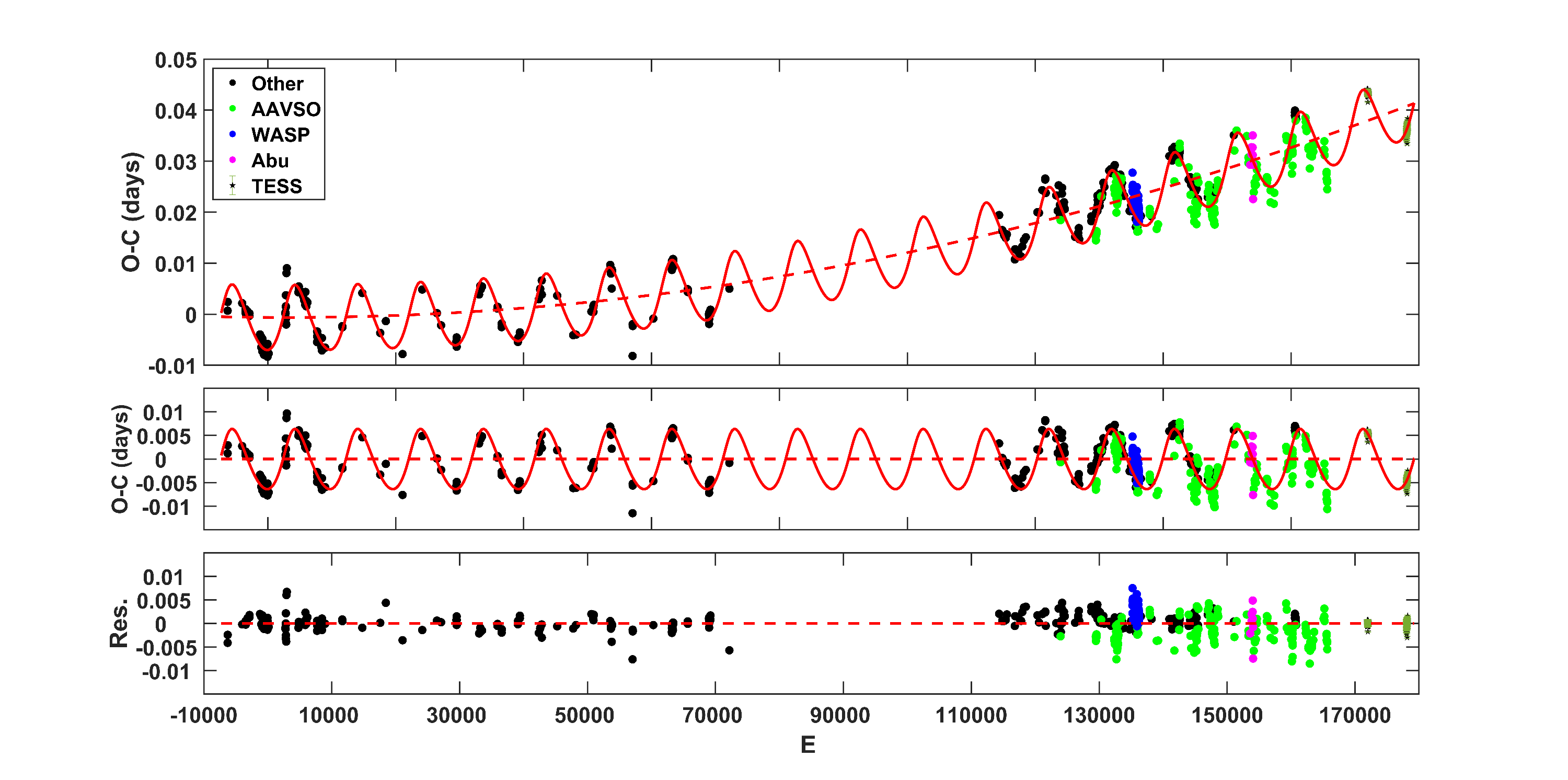}
	\caption{ The radial $O-C$ curve of SZ lyn.} \label{fig:EVT}
\end{figure}
\begin{figure}[ht!]
	\plotone{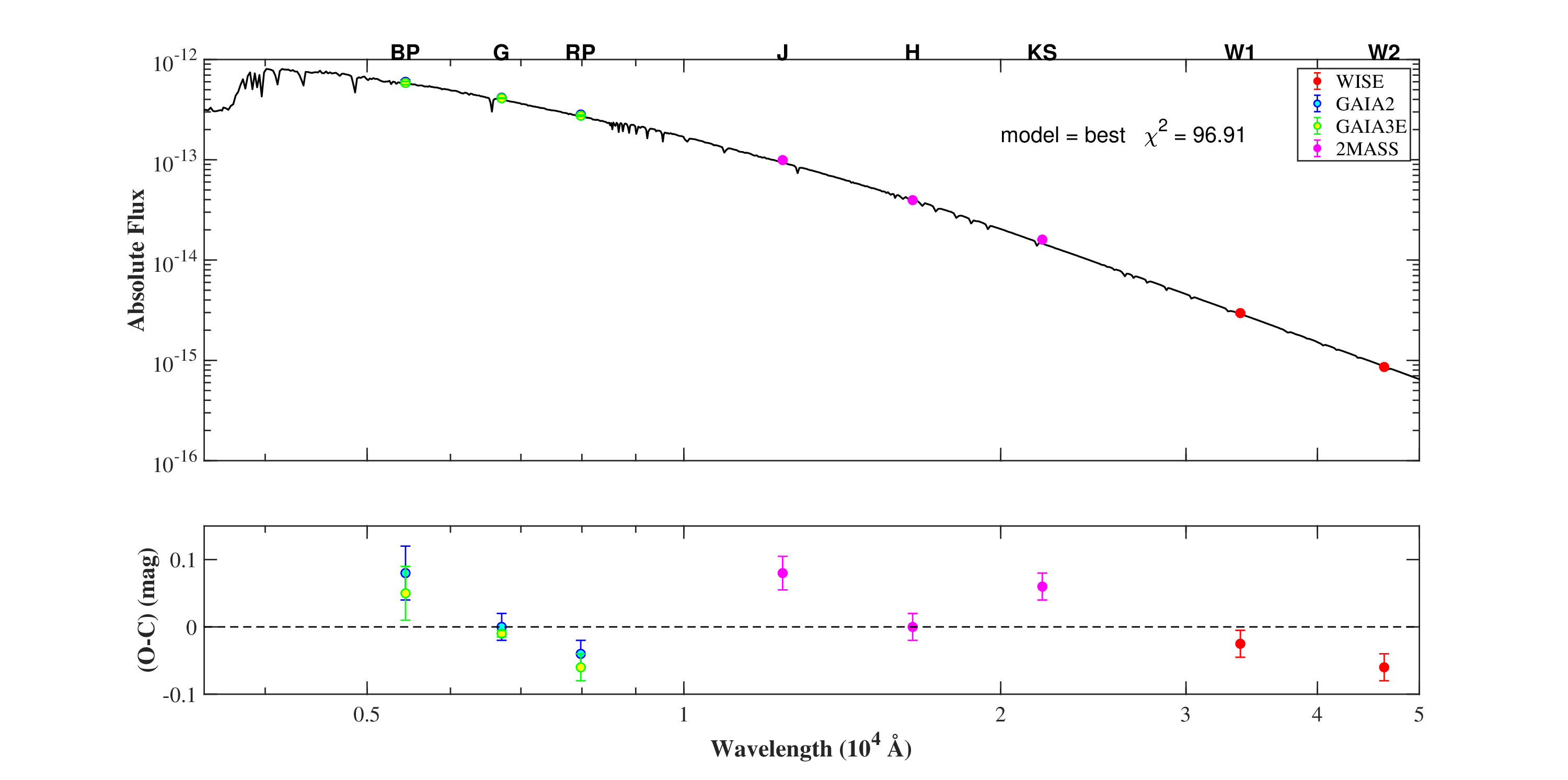}
	\caption{ The broad band spectral-energy distribution of SZ Lyn.} \label{fig:sed}
\end{figure}

We used LTTE model to fit the $O-C$ data using the MCMC method. Figure \ref{fig:EVT} shows that the fitting results in $A$=0.00651($\pm$0.00005) d, $P'$=3.24170 ($\pm$0.00075) yr, where $A$ and $P'$ are the seimi-amplitude and period of $O-C$. Therefore the mass function (lower mass limit) for SZ Lyn B is:
\begin{equation}\label{equation 7}
f(M_2)=\frac{(M_2\sin{i})^3}{(M_1+M_2)^2}=\frac{4\pi^2}{GP'^2}(a_{12} \sin{i})^3=0.1365\pm0.0031 \rm M_{\odot}
\end{equation}
where $a_{12}\sin{i} = A\times c$ (c is the speed of light) is the project seimi axis.

\subsection{ Unveiling the nature of SZ Lyn B}\label{SZ Lyn B}
The orbital inclination of SZ Lyn has been determined to be $i = 38.67 \pm 0.29^\circ$ (Figure \ref{fig:rv}) using astrometric data from Gaia, which is close to the previous value of $i = 39.6 \pm 17.7^\circ$ obtained using Hipparcos astrometric data (see  \citep{2013PASJ...65..116L}). Using Equations~\ref{equation 5} and~\ref{equation 7}, the mass of SZ Lyn B derived from radial velocity measurements is $M_2 = 1.76_{-0.042}^{+0.042} ~\mathrm{M_{\odot}}$, while $O\!-\!C$ analysis yields $M_2 = 2.07_{-0.045}^{+0.045}  ~\mathrm{M_{\odot}}$; these results are consistent within uncertainties. The corresponding orbital separations are $a_{\rm RV} = 3.342_{-0.046}^{+0.046}$ astronomical units (au) and $a_{O\!-\!C} = 3.449_{-0.035}^{+0.035}$ au, respectively. 
\begin{figure}[ht!]
	\plotone{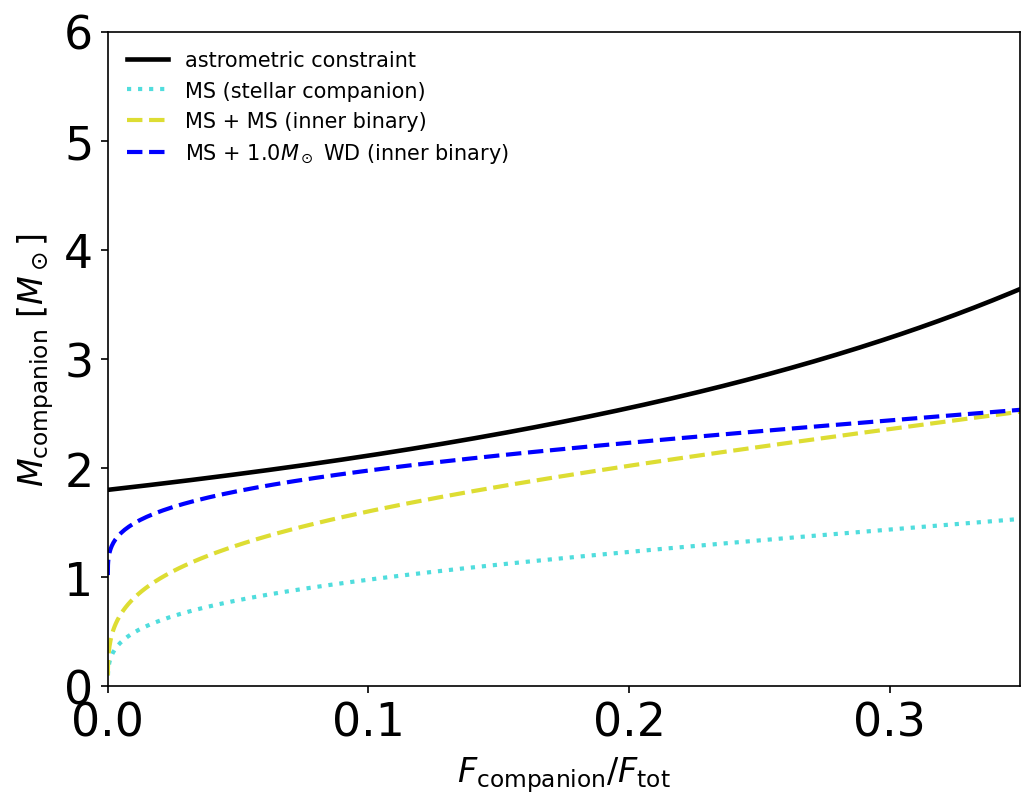}
	\caption{ Constraints on the mass of the unseen companion in SZ Lyn as a function of the $G$-band flux ratio. The solid black line shows the constraint from the astrometric mass function, setting a primary mass $M_{1}=1.83\,~\mathrm{M_{\odot}}$; a completely dark companion would imply $M_{\mathrm{2}}\approx1.76\,~\mathrm{M_{\odot}}$. If the companion contributes any light, its astrometrically inferred mass increases. The dotted cyan line shows the expected flux ratio and mass for a single main-sequence (MS) companion, which always lies below the black line, ruling out any single MS companion as the explanation for the observed orbit. The same holds for an equal-mass inner binary (yellow dashed). Even an inner binary composed of a $1.0\,~\mathrm{M_{\odot}}$ white dwarf (WD) and an MS star (blue dashed) has a mass-to-light ratio too low to match the astrometric mass function. This leaves a single NS or a WD-WD binary as the only viable companions.} \label{fig:astro_cons}
\end{figure}

Multiple lines of evidence argue against SZ~Lyn~B being a luminous companion, whether a single star or a close main-sequence binary. If the companion contributes any light (Figure~\ref{fig:astro_cons}), its astrometrically inferred mass increases. The dotted cyan line shows the expected flux ratio and mass for a single main-sequence companion, which always lies below the black line, ruling out any single main-sequence companion as the explanation for the observed orbit. The same holds for an equal-mass inner binary (yellow dashed). Even an inner binary composed of a $1.0\,~\mathrm{M_{\odot}}$ WD and a main-sequence star (blue dashed) has a mass-to-light ratio too low to match the astrometric mass function. This conclusion is further supported by: (1) the absence of spectral lines moving in antiphase to SZ~Lyn~A in both our low-resolution spectra and the high-resolution data of \cite{2024BSRSL..93..470A}; (2) the $O\!-\!C$ diagram of pulsation maxima (Figure~\ref{fig:EVT}) showing only a single variation component; and (3) the spectral energy distribution (SED) (Figure~\ref{fig:sed}) being well fitted by an A-type star model at both blue and red ends, with no excess indicative of an additional luminous component. Therefore, SZ~Lyn~B is unlikely to be any type of luminous companion.

If SZ~Lyn~B is a compact star, its mass exceeds the Chandrasekhar limit and falls within the typical range of NS masses, ruling out a WD interpretation. Moreover, spectroscopic analysis reveals no broad WD absorption features near the Balmer lines in either LAMOST (Figure~\ref{fig:spec}) or high-resolution spectra~\citep{2024BSRSL..93..470A}. A single WD companion is therefore strongly disfavoured. A black hole companion is likewise ruled out by the inferred mass, which lies well below the typical black hole mass range. A NS thus emerges as the most straightforward explanation, and we adopt this as our default assumption throughout the paper.

We therefore propose SZ~Lyn~B as a NS candidate—the most parsimonious interpretation consistent with all observational constraints. We note that the possibility of a close binary containing two low-mass WDs is not directly ruled out by current observations. However, given that SZ~Lyn~A has an age of 1.254~Gyr (Table~\ref{tab: para}) and that the WD progenitors would have had main-sequence lifetimes of at most a few Gyr, both WDs would be older than SZ~Lyn~A, and SZ~Lyn~B could not have been captured by SZ~Lyn~A. Therefore, SZ~Lyn~B is unlikely to be a close binary containing two low-mass WDs. Crucially, at periastron, the radius of SZ~Lyn~A remains well within its Roche lobe, indicating negligible mass transfer—which naturally explains the absence of detectable accretion-powered X-ray or gamma-ray emission from the NS candidate.

\subsection{The Future Prospects of SZ Lyn}
Future observations will be essential for confirming the nature of SZ~Lyn~B. Precise astrometry from Gaia~DR4 and subsequent missions can refine the orbital inclination and companion mass, while radio observations with the Five-hundred-meter Aperture Spherical Telescope (FAST) may detect potential radio signatures despite the wide separation. These efforts will help validate the neutron star interpretation and further constrain the system's evolutionary history. SZ~Lyn represents one of the first $\delta$ Scuti binaries with a NS candidate identified through the synergy of asteroseismology and dynamics, highlighting the potential of this approach to uncover non-accreting compact objects.

\section{Conclusions}
We identify SZ~Lyn as a noteworthy binary system. Asteroseismic modeling of the primary component, SZ~Lyn~A, successfully reproduces its six pulsation frequencies, yielding precise fundamental parameters: mass $M = 1.83_{-0.01}^{+0.06}~\mathrm{M_{\odot}}$, radius $R = 2.899 \pm 0.027~\mathrm{R_{\odot}}$, luminosity $L = 16.111_{-0.354}^{+0.721}~\mathrm{L_{\odot}}$, central hydrogen abundance $X_{\mathrm{c}} = 0.089_{-0.005}^{+0.032}$, and age $= 1.254_{-0.024}^{+0.079}$~Gyr. These results indicate that SZ~Lyn~A is a post-main-sequence HADS.

The companion, SZ~Lyn~B, orbits in a low-eccentricity orbit ($e=0.18$). Using the orbital inclination derived from Gaia astrometry ($i = 38.67 \pm 0.29^\circ$), we obtain companion masses of $M_2 = 1.76_{-0.042}^{+0.042}~\mathrm{M_{\odot}}$ from radial velocity analysis and $M_2 = 2.07_{-0.045}^{+0.045}~\mathrm{M_{\odot}}$ from pulsation timing variations, with corresponding orbital separations of $a_{\mathrm{RV}} = 3.342_{-0.046}^{+0.046}$~au and $a_{O-C} = 3.449_{-0.035}^{+0.035}$~au, respectively. These mass estimates place the companion well above the Chandrasekhar limit and within the typical mass range of NSs.

Multiple lines of evidence rule out luminous companions: the astrometric mass function excludes single main-sequence stars, equal-mass inner binaries, and WD--main-sequence binaries; the absence of antiphase spectral lines, a single-component $O-C$ diagram, and a spectral energy distribution fully consistent with an A-type star further support this conclusion. A WD companion is disfavored by both the mass exceeding the Chandrasekhar limit and the lack of broad Balmer absorption features, while a black hole is ruled out by the inferred mass being far below the typical black hole mass range. The possibility of a double WD binary remains observationally unconstrained but is deemed unlikely based on age arguments and capture considerations. Together, these constraints point to a NS as the most parsimonious interpretation.

This system holds significant astrophysical value. It represents a rare configuration—a wide, near-circular binary containing a pulsating star and a quiescent NS candidate—offering new constraints on supernova kick models and enabling potential studies of asteroseismology in strong gravity regimes. The discovery also demonstrates the effectiveness of combining asteroseismology with dynamical methods (radial velocity and light-travel time effect) for identifying non-accreting compact objects in wide binaries. 

\begin{acknowledgments}
This work is supported by the International Cooperation Projects of the National Key R\&D Program (No. 2022YFE0127300), the National Natural Science Foundation of China (No. 11933008), the Young Talent Project of  ``Yunnan Revitalization Talent Support Program" in Yunnan Province, the basic research project of Yunnan Province (Grant No. 202201AT070092), CAS ``Light of West China" Program. This work has made use of data from the European Space Agency (ESA) mission Gaia. (\href{https://www.cosmos.esa.int/gaia}{https: //www.cosmos.esa.int/gaia}), processed by the Gaia Data Processing and Analysis Consortium (DPAC; \href{https://www.cosmos.esa.int/web /gaia/dpac/consortium}{https://www.cosmos.esa.int/web /gaia/dpac/consortium}). Funding for the DPAC has been provided by national institutions, in particular the institutions participating in the $Gaia$ Multilateral Agreement. The $TESS$ data presented in this paper were obtained from the Mikulski Archive for Space Telescopes (MAST) at the Space Telescope Science Institute (STScI). STScI is operated by the Association of Universities for Research in Astronomy, Inc. Support to MAST for these data is provided by the NASA Office of Space Science. Funding for the $TESS$ mission is provided by the NASA Explorer Program.
\end{acknowledgments}

%

\vspace{5mm}
\facilities{TESS, LAMOST}


\software{astropy \citep{2013A&A...558A..33A,2018AJ....156..123A}, MESA \citep{2011ApJS..192....3P,2013ApJS..208....4P,2015ApJS..220...15P,2018ApJS..234...34P}, \textit{Pulse\_Adipls} \citep{2008Ap&SS.316..113C}}



\appendix

\section{Light curve data and frequency analysis}
\begin{figure}[ht!]
	\plotone{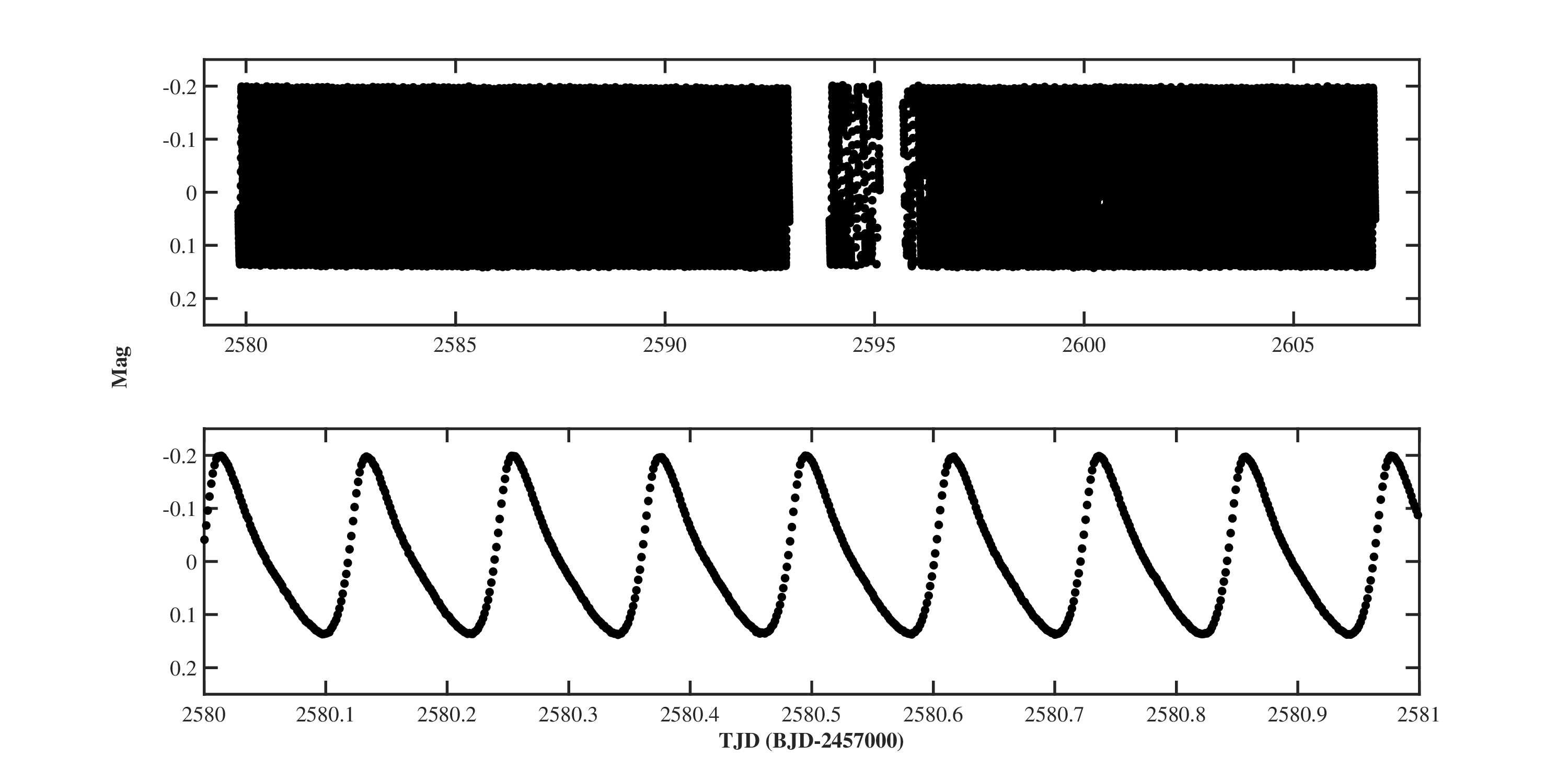}
	\caption{Light curve of SZ Lyn from TESS Sector 20. The lower panel shows a one-day zoom for clarity.\label{fig:lightcurve}}
\end{figure}

SZ Lyn was observed by the TESS mission in Sectors 20 and 47 with a 120-second cadence. We retrieved the Presearch Data Conditioning Simple Aperture Photometry (PDCSAP) flux from the Mikulski Archive for Space Telescopes (MAST) and converted it to magnitudes (Figure \ref{fig:lightcurve}).

\begin{figure}[ht!]
	\plotone{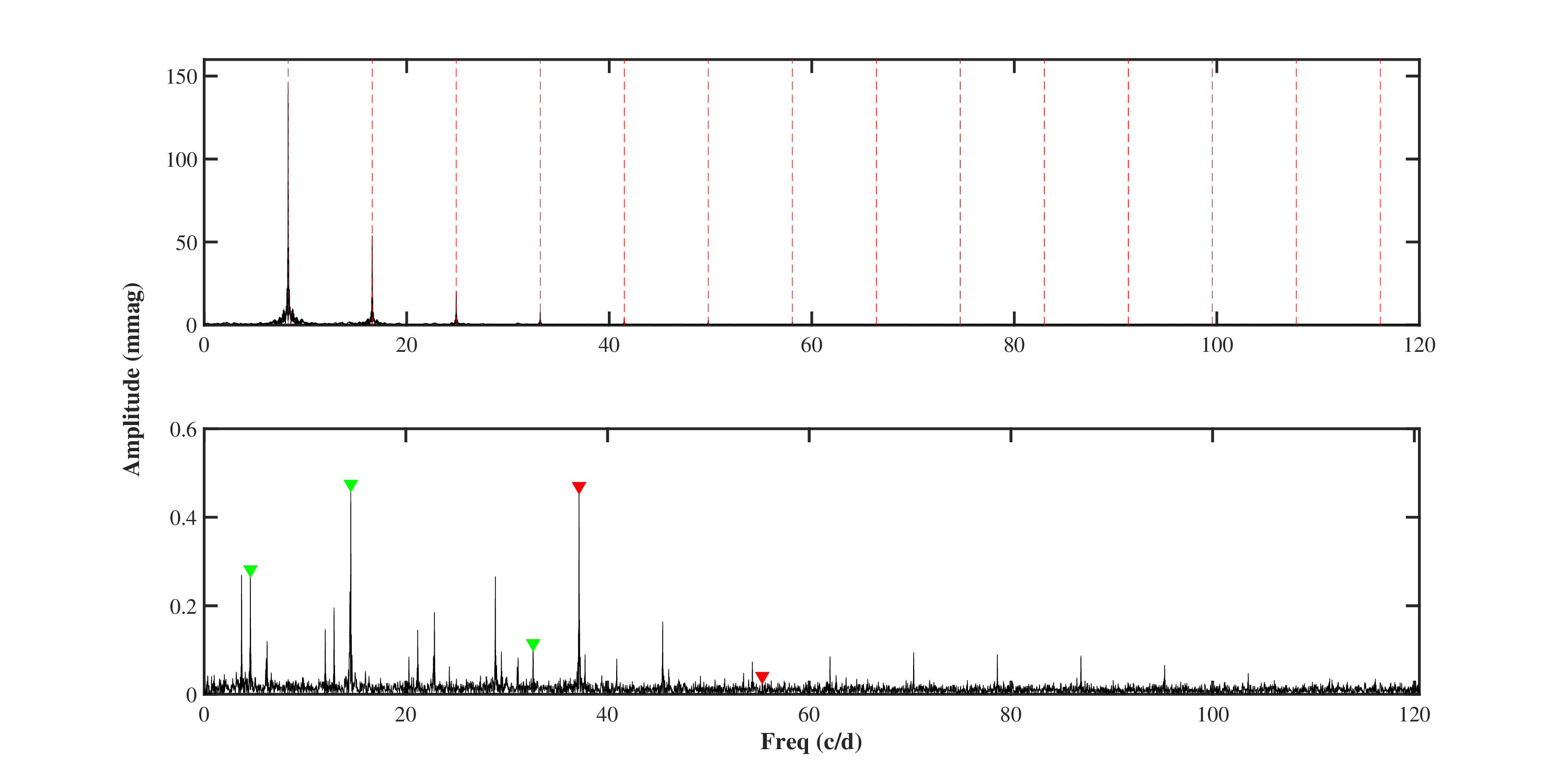}
	\caption{ Frequency spectrum of SZ Lyn. Top panel: Red dotted vertical lines (0–120 c/d) mark the fundamental mode (highest amplitude) and its harmonics. Lower panel: The six independent pulsation frequencies, listed in Table \ref{tab:freq}, with green triangles indicating non-radial modes identified by \citet{2021MNRAS.502..541A} and red triangles denoting new frequencies detected in this work.\label{fig:fourier}}
\end{figure}

To analyze the pulsation properties of SZ~Lyn, we subtracted the light-curve model from the TESS light curves. The frequency resolution for this dataset is $\delta f = 1/\Delta T \approx 0.0368$~c/d, where $\Delta T$ denotes the total observational time span covering sectors 45 and 46. We performed a detailed pulsation analysis of the light-curve residuals using multi-frequency analysis with the \textit{Period 04} software \citep
{2005CoAst.146...53L}. As no pulsation signals were detected above 120~c/d, subsequent frequency extraction was restricted to the range 0--120~c/d (Figure \ref{fig:fourier}).

At each iteration, we identified the highest-amplitude frequency and performed a multi-period least-squares fit including all previously detected frequencies. The data were then pre-whitened using this solution, and the resulting residuals formed the basis for subsequent analysis. This process iterated until residuals satisfied the signal-to-noise criterion S/N $\geq$ 4.0 \citep{1993A&A...271..482B}. Following \citet{1999DSSN...13...28M}, we computed the noise within a 1~c/d bandwidth centered on each frequency, and determined corresponding amplitudes and their uncertainties.
\begin{deluxetable*}{lccccccc}  
	\tabletypesize{\scriptsize}  
	\tablecaption{Pulsating frequencies of SZ Lyn. \label{tab:freq}}  
	\tablehead{  
		\colhead{$ID$}  & \colhead{Frequency ($c/d$)} & \colhead{Frequency ($\mu$Hz)} & \colhead{error ($c/d$)} & \colhead{Amplitude (mag)} & \colhead{error (mag)} & \colhead{SNR}
	}  
	\startdata  
	$F_1$ &  8.2961135 & 96.0198326 & 0.0000011 & 0.1453108 & 0.0000084 & 10720.80 \\  
	$F_2$ &  14.53716 & 168.25419 & 0.00034 & 0.0004918 & 0.0000084 & 54.99 \\  
	$F_3$ &  37.17183 & 430.22954 & 0.00035 & 0.0004795 & 0.0000084 & 42.06 \\  
	$F_4$ &  32.6165 & 377.5065 & 0.0015 & 0.0001110 & 0.0000084 & 7.40 \\  
	$F_5$ &  4.58526 & 53.07015 & 0.00060 & 0.0002833 & 0.0000084 & 21.62 \\  
	$F_6$ &  55.3391 & 640.4991 & 0.0039 & 0.0000436 & 0.0000084 & 5.17 \\  
	\tableline  
	\enddata  
\end{deluxetable*}

We examined the extracted frequencies to identify linear combination frequencies of the form $F_{k} = m F_{i} + n F_{j}$ \citep{2012AN....333.1053P,2015MNRAS.450.3015K}, where $m$ and $n$ are integers, $F_{i}$ and $F_{j}$ represent parent frequencies, $F_{k}$ denotes the combination frequency. A peak is classified as a combination if it satisfies two criteria: (1) both parent frequencies have amplitudes greater than the presumed combination term, and (2) the difference between observed and predicted frequencies is smaller than the frequency resolution $\delta f$ \citep{1978Ap&SS..56..285L,2019AJ....157...17L}. Finally, Six independent frequencies ($F_1$--$F_6$) were identified (Table~\ref{tab:freq}), including two tentative new frequencies ($F_3$, $F_6$) (red triangles in Figure \ref{fig:fourier}). The remaining frequencies correspond to those reported by \citet{2021MNRAS.502..541A} within our frequency resolution.

\section{Stellar Model and Asteroseismic Analysis for SZ Lyn }
\subsection{Input Physics}
We thus utilize the stellar evolution code Modules for Experiments in Stellar Astrophysics (MESA, version r22.11.1), developed by \citet{2011ApJS..192....3P,2013ApJS..208....4P,2015ApJS..220...15P,2018ApJS..234...34P}, to compute evolutionary and pulsational models for component B. Specifically, for generating stellar evolution models and calculating the adiabatic frequencies of radial and non-radial modes, we employ the \textit{pulse\_adipls} \citep{2008Ap&SS.316..113C} submodule in MESA. Our calculations adopt the 2005 update of the OPAL equation of state tables \citep{2002ApJ...576.1064R}. For opacity, we utilize the OPAL high-temperature tables from \cite{1996ApJ...464..943I} and the low-temperature tables from \citet{2005ApJ...623..585F}. We assume the initial metallicity is identical to the solar value \citep{2009ARA&A..47..481A}. Convective zones are treated using the classical mixing-length theory \citep{1958ZA.....46..108B} with a mixing-length parameter $\alpha = 1.90$ \citep{2011ApJS..192....3P}.

To model convective core overshooting, we adopt an exponentially decaying prescription for the overshooting - mixing diffusion coefficient \citep{1996A&A...313..497F, 2000A&A...360..952H}:
\begin{eqnarray}\label{equation (2)}
	\mathit{D}_{\rm ov} = \mathit{D}_0 \exp \left( \frac{-2z}{\mathit{f}_{\rm ov} \mathit{H}_{\rm p}} \right),
\end{eqnarray}
where $\mathit{D}_0$ is the diffusion coefficient at the edge of the convective core, $z$ is the distance into the radiative zone from the edge, $\mathit{f}_{\rm ov}$ is an adjustable parameter describing the efficiency of overshooting, and $\mathit{H}_{\rm p}$ is the pressure scale height. The lower limit of the diffusion coefficient is set to $\mathit{D}^{\rm limit}_{\rm ov} = 1 \times 10^{-2}$ cm$^2$ s$^{-1}$ \citep{2019ApJ...887..253C}, below which overshooting is neglected in our models. In addition, we do not include elemental diffusion, stellar rotation, and magnetic fields in the stellar structure and evolution calculations.

\subsection{Grid of Stellar Models for SZ Lyn A}
Stellar evolution paths and internal structures depend critically on the initial mass $M$, initial chemical composition $(X, Y, Z)$, and the convective overshooting parameter $\mathit{f}_{\rm ov}$. Following \citet{2018MNRAS.475..981L}, we adopt the helium abundance relation $Y = 0.249 + 1.33Z$, thereby reducing the number of independent parameters to $M$, $Z$, and $\mathit{f}_{\rm ov}$. We perform a grid search over stellar masses $M$ from $1.70$ to $2.00$ M$_{\odot}$ with a step of 0.01 M$_{\odot}$ according our results of \cite{2021MNRAS.502..541A}, and metallicities  $Z$ from $0.010$ to $0.020$ in steps of $0.001$. For the overshooting at the top of the convective core, we follow the method of \cite{2019ApJ...887..253C} to take the case of moderate overshooting ($\mathit{f}_{\rm ov} = 0.01$). 

Each star in the grid is computed from the zero-age-main-sequence to the post-main-sequence stage ($X_c = 1\times\ 10^{-5}$). The effective temperatures of $\delta$ Scuti normally vary between 6000 K and 9800 K \citep{2010aste.book.....A}. we restrict the effective temperature of stellar models  inside this range. Luminosity and radius are constrained to $1.24 < \log (L/$L$_{\odot}) < 1.53$ and $0.39 <  \log (R/$R$_{\odot}) < 0.53 $, based on results from our light curve modeling.  As a star moves along its evolutionary track into this region, the adiabatic frequencies of the radial ($\ell$ = 0) and non-radial ($\ell$ = 1, 2, and 3) oscillations are calculated for the structure model at each stage of its evolution.

Additionally, we also include stellar rotation as a fourth adjustable parameter, varying the rotation velocity $\mathit{V}_{\rm rot}$ from 1  to 20 km s$^{-1}$ in steps of 1 km s$^{-1}$ due to the component B's rotation velocity is $v_{rot} = 10$ km s$^{-1}$ \citep{2024BSRSL..93..470A}. For each pulsation mode at a given $\mathit{P}_{\rm rot}$, rotational splitting produces $2l + 1$ frequency components according to:
\begin{eqnarray}\label{equation (3)}
	\nu_{l,n,m} = \nu_{l,n} + m\delta\nu_{l,n} = \nu_{l,n} + \beta_{l,n} \frac{m}{\mathit{P}_{\rm rot}}
\end{eqnarray}
\citep{1981ApJ...244..299S,1992ApJ...394..670D,2010aste.book.....A}, where $\delta\nu_{l,n}$ is the rotational splitting frequency and $\beta_{l,n}$ determines the splitting magnitude. For uniformly rotating stars, $\beta_{l,n}$ is given by \cite{2010aste.book.....A}:
\begin{eqnarray}\label{equation (4)}
	\beta_{l,n} = \frac{\int_{0}^{R} \left( \xi^2_{r} + L^2\xi^2_{h} - 2\xi_{r}\xi_{h} - \xi^2_{h} \right) r^2 \rho  \,dr}{\int_{0}^{R} \left( \xi^2_{r} + L^2\xi^2_{h} \right) r^2 \rho \,dr},
\end{eqnarray}
where $\xi_{r}$ and $\xi_{h}$ are the radial and horizontal displacement eigenfunctions, $\rho$ is the local density, and $L^2 = l(l+1)$. According to Equation \ref{equation (3)}, each dipole mode splits into three different components, corresponding to modes with $m$ = $-$1, 0, and $+$1, respectively. Each quadrupole mode splits into five different components, corresponding to modes with $m$ = $-$2, $-$1, 0, $+$1, and $+$2, respectively.

\subsection{Fitting Results of SZ Lyn B}
We compare model frequencies with spherical degrees $l = 0$, $1$, $2$, and $3$ to the observed frequencies $F_1$, $F_2$, $F_3$, $F_4$, $F_5$, and $F_6$ to identify the optimal model. The goodness of fit is evaluated using:
\begin{eqnarray}\label{equation (5)}
	S^2 = \frac{1}{k} \sum_{i=1}^{k} |f_{\mathrm{model},i} - f_{\mathrm{obs},i}|^2,
\end{eqnarray}
where $f_{\mathrm{obs},i}$ is the $i$-th observed frequency, $f_{\mathrm{model},i}$ is the corresponding model frequency, and $k$ is the number of observed modes. The frequencies $F_3$ and $F_6$ could not be precisely identified in advance; therefore, the nearest theoretical frequencies were assigned to represent them in the model.
\begin{deluxetable*}{lcccc}  
	\tabletypesize{\scriptsize}  
	\tablecaption{Comparisons between model frequencies of best-fitting model (Model A55) and observations.\label{tab: best_fit}}  
	\tablehead{  
		\colhead{ID} & \colhead{$F_{\rm mod}$ ($\mu$Hz)} & \colhead{$F_{\rm obs}$ ($\mu$Hz)} & \colhead{$(\ell,\,n_p,\,n_g,\,m)$} & \colhead{$|F_{\rm mod} - F_{\rm obs}|$ ($\mu$Hz)}  
	}  
	\startdata  
	$F_1$  & 96.2190 & 96.0185 & $(0,\,0,\,0,\,0)$   & 0.2005 \\  
	$F_2$  & 168.3562 & 168.2291 & $(2,\,-2,\,2,\,-1)$  & 0.1271 \\ 
	$F_3$&  430.4194 & 430.2295 & $(1,\,11,\,0,\,-1)$   & 0.1899 \\
	$F_4$  & 377.4836 & 377.5462 & $(2,\,8,\,0,\,1)$   & 0.0626\\  
	$F_5$  & 53.0423 & 53.0555 & $(3,\,0,\,-14,\,-3)$  & 0.0132 \\  
	$F_6$& 640.5245 & 640.4991 & $(3,\,16,\,0,\,0)$   & 0.0254 \\
	\tableline  
	\enddata  
	\tablecomments{$F_{\rm obs}$ is the observed frequency. $F_{\rm mod}$ is the model frequency. ($\ell$, $n_p$, $n_g$, $m$) are the spherical harmonic degree, the radial orders in
			the p-mode propagation zone, the radial orders in the g-mode propagation zone, and the azimuthal number of themodel frequency.}
\end{deluxetable*}
\begin{deluxetable*}{lcc}  
	\tabletypesize{\scriptsize}  
	\tablecaption{ Fundamental parameters of the component A of SZ Lyn.\label{tab: para}}
	\tablehead{  
		\colhead{Parameters} & \colhead{Single-star Models}
	}
	\startdata  
	$Z$ & $0.015$--$0.018$ ($0.015^{+0.003}_{-0.000}$)  \\
	$M$ (M$_\odot$) & $1.82$--$1.89$ ($1.83^{+0.06}_{-0.01}$) \\
	$V_{\rm rot}$ (kms$^{-1}$) & $2$--$20$ ($9^{+11}_{-7}$)\\
	$T_{\rm eff}$ (K) & $6732$--$6842$ ($6791^{+51}_{-58}$)\\
	$\log(g)$ (cms$^{-2}$) & $3.774$--$3.782$ ($3.776^{+0.006}_{-0.002}$) \\
	$R$ (R$_\odot$) & $2.899$--$2.906$ ($2.899^{+0.027}_{-0.000}$) \\
	$L$ (L$_\odot$) & $15.758$--$16.833$ ($16.112^{+0.721}_{-0.354}$) \\
	$\tau_0$ (hr) & $3.748$--$3.759$ ($3.750^{+0.009}_{-0.002}$) \\
	$X_c$ & $0.084$--$0.121$ ($0.089^{+0.032}_{-0.005}$)\\
	$Age$ (Gyr) & $1.230$--$1.333$ ($1.254^{+0.079}_{-0.024}$) \\
	\tableline  
	\enddata  
	
	\tablecomments{$V_{\rm rot}$ denotes the rotation velocity. $\tau_0$ is the acoustic radius. $X_c$ is the central hydrogen abundance.}
	
\end{deluxetable*}
\begin{figure}[ht!]
	\plotone{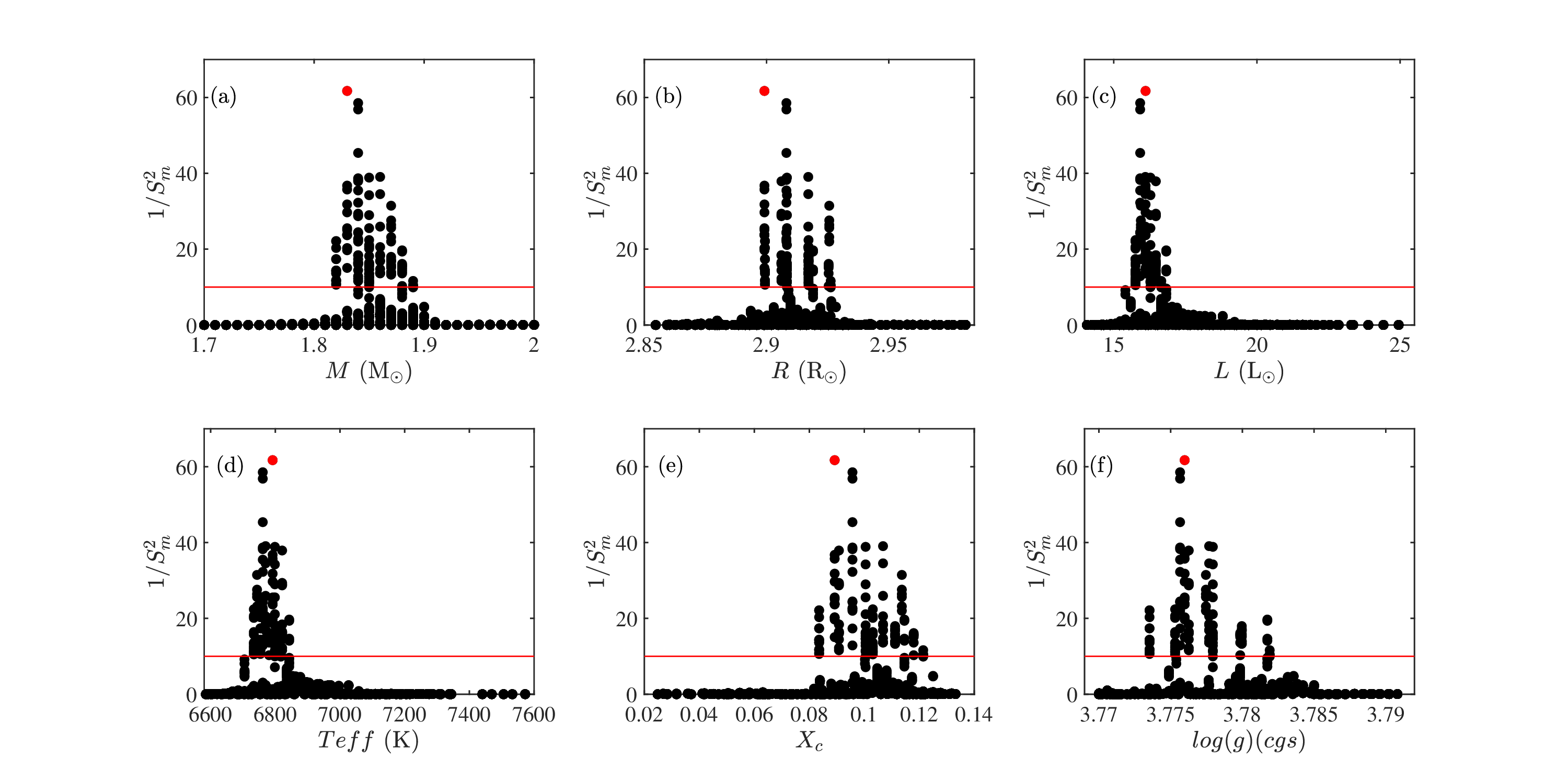}
	\caption{This visualization presents the fitting metric $S^{2}_m$ and physical parameters: stellar mass ($M$), radius ($R$), luminosity ($L$), effective temperature ($T\mathrm{eff}$), central hydrogen abundance ($X_c$), and gravitational acceleration ($\log g$) with a red horizontal line marking the threshold $S^{2}_m = 0.10$, where black dots indicate the minimum $S^{2}_m$ for each model while the red dot identifies the minimum $S^{2}_m$ of the best-fitting model.\label{fig:MESA_1}}
\end{figure}
\begin{figure}[ht!]
	\plotone{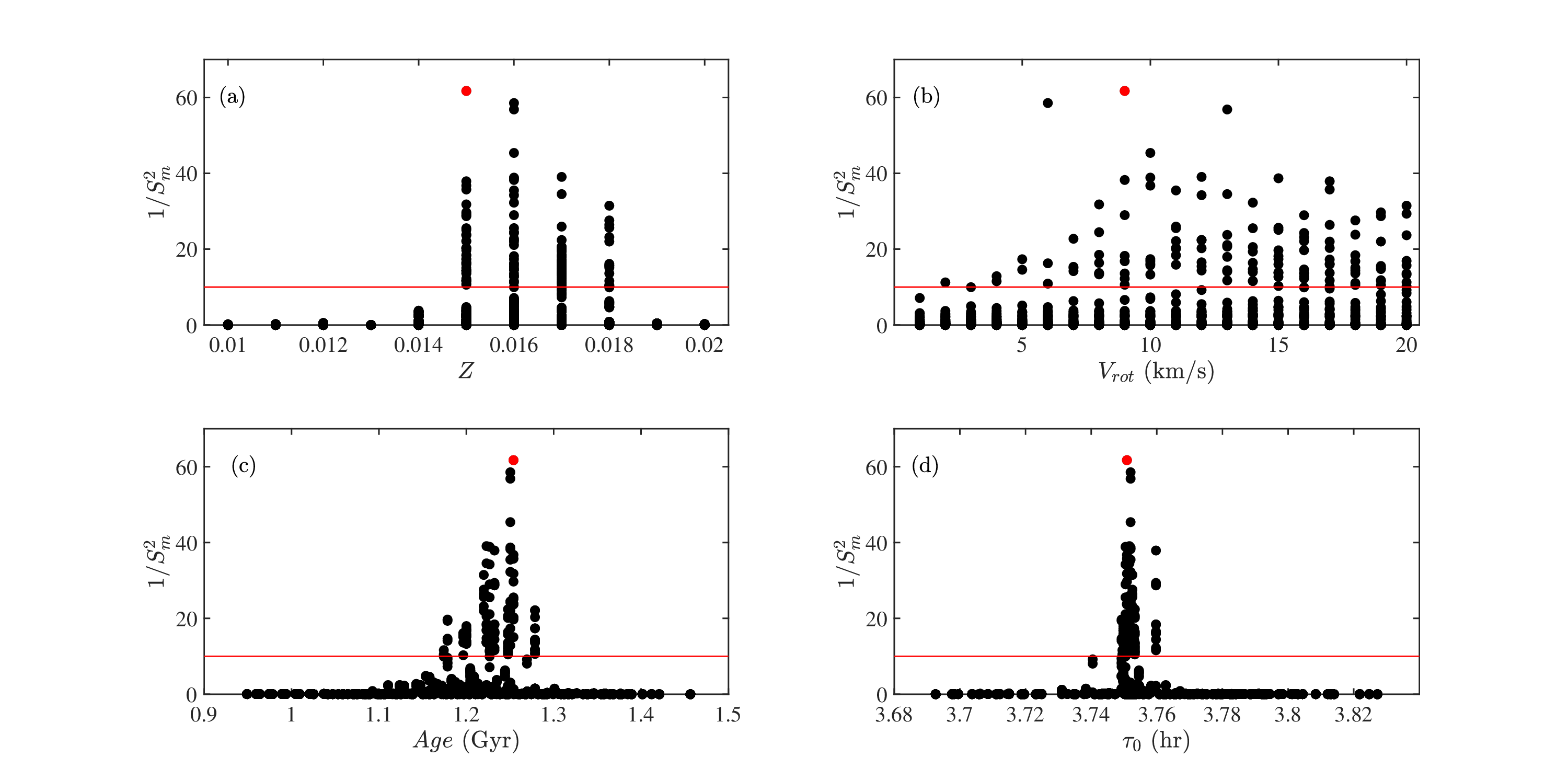}
	\caption{This visualization presents the fitting results $S^{2}_m$ alongside key physical parameters: metallicity ($Z$), stellar rotation velocity ($V_{\rm rot}$), stellar age ($Age$), and accoustic radius ($\tau_0$). A red horizontal line marks the threshold $S^{2}_m = 0.10$. Black dots denote the minimum $S^{2}_m$ for each model, while a red dot identifies the minimum $S^{2}_m$ for the best-fitting model.
		\label{fig:MESA_2}}
\end{figure}

Figures \ref{fig:MESA_1} and \ref{fig:MESA_2} display the variation of the goodness-of-fit $S^2_{m}$ across different physical parameters.  Each dot in the figures represents one minimum value of $S^2$ along one evolutionary track, denoted as $S^2_{m}$. Horizontal lines in both figures indicate $S^2_{m} = 0.10$, corresponding to the squared value of the frequency resolution $\delta f$. Dots above this threshold correspond to 84 candidate models that are listed in Table \ref{tab:model_para}. Model A55 achieves the global minimum $S^2_{m} = 0.0162$, establishing it as the optimal solution. This best-fitting model is highlighted by red dots in Figures \ref{fig:MESA_1} and \ref{fig:MESA_2}.

Figures \ref{fig:MESA_1} (a)--(c) display the variation of $S^2_2$ versus stellar mass ($M$), radius ($R$), and luminosity ($L$). These parameters demonstrate excellent convergence with best-fit values $M = 1.83^{+0.06}_{-0.01}$ M$_{\odot}$, $R = 2.899^{+0.027}_{-0.000}$ R$_{\odot}$, and $L = 16.111^{+0.721}_{-0.354}$ L$_{\odot}$. Meanwhile, Figures \ref{fig:MESA_1} (d)--(f) show $S^2_2$ as functions of effective temperature ($T\mathrm{eff}$), gravitational acceleration ($\log g$), and central hydrogen abundance ($X_{\mathrm{c}}$). The parameters effective temperature and gravitational acceleration exhibit tight constraints: $T\mathrm{eff} = 6791^{+51}_{-58}~\mathrm{Gyr}$ and $\log g = 3.776^{+0.006}_{-0.002}~\mathrm{cm\,s^{-2}}$. In contrast, the central hydrogen abundance spans a broader range $X_{\mathrm{c}} = 0.089^{+0.032}_{-0.005}$.

Figures \ref{fig:MESA_2} (a)--(d) depict the variation of $S^2_2$ with metallicity ($Z$), rotation velocity ($V_{\mathrm{rot}}$), stellar age ($\mathrm{Age}$) and acoustic radius ($\tau_0$). Excellent convergence is observed for metallicity ($Z$), stellar age ($Age$) and acoustic radius ($\tau_0$) with best-fit values $Z = 0.016^{+0.008}_{-0.001}$, $\mathrm{Age} = 1.254^{+0.079}_{-0.024}~\mathrm{Gyr}$ and $\tau_0 = 3.750^{+0.009}_{-0.002}~\mathrm{hr}$, while the stellar rotation shows a broader distribution $T_{\mathrm{eff}} = 9^{+11}_{-7}~\mathrm{km/s}$. 

Table~\ref{tab: best_fit} presents a comparison between the observed frequencies and those of the best-fitting model (Model A55). Based on this comparison, new frequencies $F_{5}$ and $F_6$ are identified as two non radial modes with $l$=1 and 3, respectively. $F_1$ is classified as a radial mode, while $F_2$, $F_3$, and $F_4$ are recognized as non radial modes with $l$=2 and 3 that is matched with the results of mode identification in \cite{2021MNRAS.502..541A}. The asteroseismically derived parameters of component~A in SZ Lyn are therefore listed in Column~2 of Table~\ref{tab: para}.

\setlength{\tabcolsep}{3pt}
\renewcommand{\arraystretch}{0.9}
\footnotesize 
\begin{longtable}{lccccccccccccc}
	\caption{Candidate models with $S^2_m < 0.10$.\label{tab:model_para}} \\
	\hline
	\multicolumn{1}{c}{Mode ID} & \multicolumn{1}{c}{$X_c$} & \multicolumn{1}{c}{$M$ (M$_\odot$)} & \multicolumn{1}{c}{$T_{\rm eff}$ (K)} & \multicolumn{1}{c}{log $g$ (cms$^{-2}$)} & \multicolumn{1}{c}{$R$ (R$_\odot$)} & \multicolumn{1}{c}{$L$ (L$_\odot$)} & \multicolumn{1}{c}{Age (Gyr)} & \multicolumn{1}{c}{$Z$} & \multicolumn{1}{c}{$V_{\rm rot}$ (km/s)} & \multicolumn{1}{c}{$\tau_0$ (hr)} & \multicolumn{1}{c}{$S^2_m$} \\
	\hline
	\endfirsthead
	
	\multicolumn{13}{c}%
	{{\bfseries \tablename\ \thetable{} -- continued}} \\
	\hline
	\multicolumn{1}{c}{Mode ID} & \multicolumn{1}{c}{$X_c$} & \multicolumn{1}{c}{$M$ (M$_\odot$)}& \multicolumn{1}{c}{$T_{\rm eff}$ (K)}  & \multicolumn{1}{c}{log $g$ (cms$^{-2}$)} & \multicolumn{1}{c}{$R$ (R$_\odot$)} & \multicolumn{1}{c}{$L$ (L$_\odot$)} & \multicolumn{1}{c}{Age (Gyr)} & \multicolumn{1}{c}{$Z$} & \multicolumn{1}{c}{$V_{\rm rot}$ (km/s)} & \multicolumn{1}{c}{$\tau_0$ (hr)} & \multicolumn{1}{c}{$S^2_m$} \\
	\hline
	\endhead
	
	\hline \multicolumn{13}{r}{{Continued}} \\
	\endfoot
	
	\hline
	\multicolumn{13}{l}{{\small $V_{\rm rot}$ is the rotation velocity}. $\tau_0$ is the acoustic radius. $X_c$ is the central hydrogen abundance.}\\
	\endlastfoot
A1 & 0.0957 & 1.84 & 6761 & 3.776 & 2.908 & 15.922 & 1.250 & 0.0160 & 7 & 3.752 & 0.0440 \\ 
A2 & 0.0957 & 1.84 & 6761 & 3.776 & 2.908 & 15.922 & 1.250 & 0.0160 & 15 & 3.752 & 0.0258 \\ 
A3 & 0.0957 & 1.84 & 6761 & 3.776 & 2.908 & 15.922 & 1.250 & 0.0160 & 14 & 3.752 & 0.0310 \\ 
A4 & 0.0957 & 1.84 & 6761 & 3.776 & 2.908 & 15.922 & 1.250 & 0.0160 & 16 & 3.752 & 0.0412 \\ 
A5 & 0.0957 & 1.84 & 6761 & 3.776 & 2.908 & 15.922 & 1.250 & 0.0160 & 10 & 3.752 & 0.0220 \\ 
A6 & 0.0957 & 1.84 & 6761 & 3.776 & 2.908 & 15.922 & 1.250 & 0.0160 & 6 & 3.752 & 0.0171 \\ 
A7 & 0.0957 & 1.84 & 6761 & 3.776 & 2.908 & 15.922 & 1.250 & 0.0160 & 9 & 3.752 & 0.0261 \\ 
A8 & 0.0957 & 1.84 & 6761 & 3.776 & 2.908 & 15.922 & 1.250 & 0.0160 & 8 & 3.752 & 0.0409 \\ 
A9 & 0.0957 & 1.84 & 6761 & 3.776 & 2.908 & 15.922 & 1.250 & 0.0160 & 11 & 3.752 & 0.0282 \\ 
A10 & 0.0957 & 1.84 & 6761 & 3.776 & 2.908 & 15.922 & 1.250 & 0.0160 & 4 & 3.752 & 0.0777 \\ 
A11 & 0.0957 & 1.84 & 6761 & 3.776 & 2.908 & 15.922 & 1.250 & 0.0160 & 17 & 3.752 & 0.0455 \\ 
A12 & 0.0957 & 1.84 & 6761 & 3.776 & 2.908 & 15.922 & 1.250 & 0.0160 & 12 & 3.752 & 0.0446 \\ 
A13 & 0.0957 & 1.84 & 6761 & 3.776 & 2.908 & 15.922 & 1.250 & 0.0160 & 13 & 3.752 & 0.0176 \\ 
A14 & 0.0957 & 1.84 & 6761 & 3.776 & 2.908 & 15.922 & 1.250 & 0.0160 & 5 & 3.752 & 0.0577 \\ 
A15 & 0.0836 & 1.82 & 6754 & 3.774 & 2.899 & 15.757 & 1.279 & 0.0150 & 13 & 3.753 & 0.0850 \\ 
A16 & 0.0836 & 1.82 & 6754 & 3.774 & 2.899 & 15.757 & 1.279 & 0.0150 & 8 & 3.753 & 0.0730 \\ 
A17 & 0.0836 & 1.82 & 6754 & 3.774 & 2.899 & 15.757 & 1.279 & 0.0150 & 16 & 3.753 & 0.0452 \\ 
A18 & 0.0836 & 1.82 & 6754 & 3.774 & 2.899 & 15.757 & 1.279 & 0.0150 & 18 & 3.753 & 0.0696 \\ 
A19 & 0.0836 & 1.82 & 6754 & 3.774 & 2.899 & 15.757 & 1.279 & 0.0150 & 19 & 3.753 & 0.0846 \\ 
A20 & 0.0836 & 1.82 & 6754 & 3.774 & 2.899 & 15.757 & 1.279 & 0.0150 & 11 & 3.753 & 0.0452 \\ 
A21 & 0.0836 & 1.82 & 6754 & 3.774 & 2.899 & 15.757 & 1.279 & 0.0150 & 6 & 3.753 & 0.0917 \\ 
A22 & 0.0836 & 1.82 & 6754 & 3.774 & 2.899 & 15.757 & 1.279 & 0.0150 & 7 & 3.753 & 0.0704 \\ 
A23 & 0.0836 & 1.82 & 6754 & 3.774 & 2.899 & 15.757 & 1.279 & 0.0150 & 12 & 3.753 & 0.0696 \\ 
A24 & 0.0836 & 1.82 & 6754 & 3.774 & 2.899 & 15.757 & 1.279 & 0.0150 & 15 & 3.753 & 0.0576 \\ 
A25 & 0.0836 & 1.82 & 6754 & 3.774 & 2.899 & 15.757 & 1.279 & 0.0150 & 14 & 3.753 & 0.0863 \\ 
A26 & 0.0836 & 1.82 & 6754 & 3.774 & 2.899 & 15.757 & 1.279 & 0.0150 & 10 & 3.753 & 0.0576 \\ 
A27 & 0.0836 & 1.82 & 6754 & 3.774 & 2.899 & 15.757 & 1.279 & 0.0150 & 9 & 3.753 & 0.0942 \\ 
A28 & 0.0836 & 1.82 & 6754 & 3.774 & 2.899 & 15.757 & 1.279 & 0.0150 & 17 & 3.753 & 0.0492 \\ 
A29 & 0.0836 & 1.82 & 6754 & 3.774 & 2.899 & 15.757 & 1.279 & 0.0150 & 20 & 3.753 & 0.0736 \\ 
A30 & 0.1005 & 1.85 & 6799 & 3.778 & 2.908 & 16.281 & 1.227 & 0.0160 & 12 & 3.750 & 0.0292 \\ 
A31 & 0.1005 & 1.85 & 6799 & 3.778 & 2.908 & 16.281 & 1.227 & 0.0160 & 11 & 3.750 & 0.0392 \\ 
A32 & 0.1005 & 1.85 & 6799 & 3.778 & 2.908 & 16.281 & 1.227 & 0.0160 & 5 & 3.750 & 0.0686 \\ 
A33 & 0.1005 & 1.85 & 6799 & 3.778 & 2.908 & 16.281 & 1.227 & 0.0160 & 4 & 3.750 & 0.0867 \\ 
A34 & 0.1005 & 1.85 & 6799 & 3.778 & 2.908 & 16.281 & 1.227 & 0.0160 & 6 & 3.750 & 0.0614 \\ 
A35 & 0.1005 & 1.85 & 6799 & 3.778 & 2.908 & 16.281 & 1.227 & 0.0160 & 7 & 3.750 & 0.0652 \\ 
A36 & 0.1005 & 1.85 & 6799 & 3.778 & 2.908 & 16.281 & 1.227 & 0.0160 & 2 & 3.750 & 0.0890 \\ 
A37 & 0.1005 & 1.85 & 6799 & 3.778 & 2.908 & 16.281 & 1.227 & 0.0160 & 8 & 3.750 & 0.0738 \\ 
A38 & 0.1005 & 1.85 & 6799 & 3.778 & 2.908 & 16.281 & 1.227 & 0.0160 & 9 & 3.750 & 0.0345 \\ 
A39 & 0.1005 & 1.85 & 6799 & 3.778 & 2.908 & 16.281 & 1.227 & 0.0160 & 10 & 3.750 & 0.0257 \\ 
A40 & 0.1005 & 1.85 & 6799 & 3.778 & 2.908 & 16.281 & 1.227 & 0.0160 & 13 & 3.750 & 0.0474 \\ 
A41 & 0.1005 & 1.85 & 6799 & 3.778 & 2.908 & 16.281 & 1.227 & 0.0160 & 16 & 3.750 & 0.0781 \\ 
A42 & 0.1005 & 1.85 & 6799 & 3.778 & 2.908 & 16.281 & 1.227 & 0.0160 & 14 & 3.750 & 0.0681 \\ 
A43 & 0.1005 & 1.85 & 6799 & 3.778 & 2.908 & 16.281 & 1.227 & 0.0160 & 15 & 3.750 & 0.0551 \\ 
A44 & 0.1005 & 1.85 & 6799 & 3.778 & 2.908 & 16.281 & 1.227 & 0.0160 & 17 & 3.750 & 0.0626 \\ 
A45 & 0.0892 & 1.83 & 6792 & 3.776 & 2.899 & 16.112 & 1.254 & 0.0150 & 11 & 3.751 & 0.0493 \\ 
A46 & 0.0892 & 1.83 & 6792 & 3.776 & 2.899 & 16.112 & 1.254 & 0.0150 & 13 & 3.751 & 0.0421 \\ 
A47 & 0.0892 & 1.83 & 6792 & 3.776 & 2.899 & 16.112 & 1.254 & 0.0150 & 10 & 3.751 & 0.0272 \\ 
A48 & 0.0892 & 1.83 & 6792 & 3.776 & 2.899 & 16.112 & 1.254 & 0.0150 & 15 & 3.751 & 0.0399 \\ 
A49 & 0.0892 & 1.83 & 6792 & 3.776 & 2.899 & 16.112 & 1.254 & 0.0150 & 19 & 3.751 & 0.0337 \\ 
A50 & 0.0892 & 1.83 & 6792 & 3.776 & 2.899 & 16.112 & 1.254 & 0.0150 & 18 & 3.751 & 0.0420 \\ 
A51 & 0.0892 & 1.83 & 6792 & 3.776 & 2.899 & 16.112 & 1.254 & 0.0150 & 16 & 3.751 & 0.0507 \\ 
A52 & 0.0892 & 1.83 & 6792 & 3.776 & 2.899 & 16.112 & 1.254 & 0.0150 & 17 & 3.751 & 0.0280 \\ 
A53 & 0.0892 & 1.83 & 6792 & 3.776 & 2.899 & 16.112 & 1.254 & 0.0150 & 8 & 3.751 & 0.0315 \\ 
A54 & 0.0892 & 1.83 & 6792 & 3.776 & 2.899 & 16.112 & 1.254 & 0.0150 & 20 & 3.751 & 0.0423 \\ 
\textbf{A55} & \textbf{0.0892} & \textbf{1.83} & \textbf{6792} & \textbf{3.776} & \textbf{2.899} & \textbf{16.112} & \textbf{1.254} & \textbf{0.0150} & \textbf{9} & \textbf{3.751} & \textbf{0.0162} \\ 
A56 & 0.0892 & 1.83 & 6792 & 3.776 & 2.899 & 16.112 & 1.254 & 0.0150 & 12 & 3.751 & 0.0494 \\ 
A57 & 0.0892 & 1.83 & 6792 & 3.776 & 2.899 & 16.112 & 1.254 & 0.0150 & 7 & 3.751 & 0.0664 \\ 
A58 & 0.0892 & 1.83 & 6792 & 3.776 & 2.899 & 16.112 & 1.254 & 0.0150 & 14 & 3.751 & 0.0392 \\ 
A59 & 0.0908 & 1.84 & 6821 & 3.776 & 2.906 & 16.471 & 1.232 & 0.0150 & 15 & 3.760 & 0.0625 \\ 
A60 & 0.0908 & 1.84 & 6821 & 3.776 & 2.906 & 16.471 & 1.232 & 0.0150 & 18 & 3.760 & 0.0543 \\ 
A61 & 0.0908 & 1.84 & 6821 & 3.776 & 2.906 & 16.471 & 1.232 & 0.0150 & 13 & 3.760 & 0.0693 \\ 
A62 & 0.0908 & 1.84 & 6821 & 3.776 & 2.906 & 16.471 & 1.232 & 0.0150 & 19 & 3.760 & 0.0348 \\ 
A63 & 0.0908 & 1.84 & 6821 & 3.776 & 2.906 & 16.471 & 1.232 & 0.0150 & 20 & 3.760 & 0.0341 \\ 
A64 & 0.0908 & 1.84 & 6821 & 3.776 & 2.906 & 16.471 & 1.232 & 0.0150 & 12 & 3.760 & 0.0605 \\ 
A65 & 0.0908 & 1.84 & 6821 & 3.776 & 2.906 & 16.471 & 1.232 & 0.0150 & 11 & 3.760 & 0.0544 \\ 
A66 & 0.0908 & 1.84 & 6821 & 3.776 & 2.906 & 16.471 & 1.232 & 0.0150 & 14 & 3.760 & 0.0863 \\ 
A67 & 0.0908 & 1.84 & 6821 & 3.776 & 2.906 & 16.471 & 1.232 & 0.0150 & 9 & 3.760 & 0.0818 \\ 
A68 & 0.0908 & 1.84 & 6821 & 3.776 & 2.906 & 16.471 & 1.232 & 0.0150 & 10 & 3.760 & 0.0618 \\ 
A69 & 0.0908 & 1.84 & 6821 & 3.776 & 2.906 & 16.471 & 1.232 & 0.0150 & 16 & 3.760 & 0.0345 \\ 
A70 & 0.0908 & 1.84 & 6821 & 3.776 & 2.906 & 16.471 & 1.232 & 0.0150 & 17 & 3.760 & 0.0264 \\ 
A71 & 0.0908 & 1.84 & 6821 & 3.776 & 2.906 & 16.471 & 1.232 & 0.0150 & 8 & 3.760 & 0.0611 \\ 
A72 & 0.1068 & 1.86 & 6770 & 3.778 & 2.917 & 16.103 & 1.223 & 0.0170 & 19 & 3.752 & 0.0678 \\ 
A73 & 0.1068 & 1.86 & 6770 & 3.778 & 2.917 & 16.103 & 1.223 & 0.0170 & 17 & 3.752 & 0.0728 \\ 
A74 & 0.1068 & 1.86 & 6770 & 3.778 & 2.917 & 16.103 & 1.223 & 0.0170 & 8 & 3.752 & 0.0540 \\ 
A75 & 0.1068 & 1.86 & 6770 & 3.778 & 2.917 & 16.103 & 1.223 & 0.0170 & 20 & 3.752 & 0.0591 \\ 
A76 & 0.1068 & 1.86 & 6770 & 3.778 & 2.917 & 16.103 & 1.223 & 0.0170 & 16 & 3.752 & 0.0732 \\ 
A77 & 0.1068 & 1.86 & 6770 & 3.778 & 2.917 & 16.103 & 1.223 & 0.0170 & 15 & 3.752 & 0.0729 \\ 
A78 & 0.1068 & 1.86 & 6770 & 3.778 & 2.917 & 16.103 & 1.223 & 0.0170 & 10 & 3.752 & 0.0594 \\ 
A79 & 0.1068 & 1.86 & 6770 & 3.778 & 2.917 & 16.103 & 1.223 & 0.0170 & 18 & 3.752 & 0.0735 \\ 
A80 & 0.1068 & 1.86 & 6770 & 3.778 & 2.917 & 16.103 & 1.223 & 0.0170 & 14 & 3.752 & 0.0486 \\ 
A81 & 0.1068 & 1.86 & 6770 & 3.778 & 2.917 & 16.103 & 1.223 & 0.0170 & 13 & 3.752 & 0.0290 \\ 
A82 & 0.1068 & 1.86 & 6770 & 3.778 & 2.917 & 16.103 & 1.223 & 0.0170 & 11 & 3.752 & 0.0385 \\ 
A83 & 0.1068 & 1.86 & 6770 & 3.778 & 2.917 & 16.103 & 1.223 & 0.0170 & 12 & 3.752 & 0.0256 \\ 
A84 & 0.1068 & 1.86 & 6770 & 3.778 & 2.917 & 16.103 & 1.223 & 0.0170 & 9 & 3.752 & 0.0548 \\ 
A85 & 0.1031 & 1.85 & 6734 & 3.775 & 2.917 & 15.761 & 1.248 & 0.0170 & 8 & 3.753 & 0.0749 \\ 
A86 & 0.1031 & 1.85 & 6734 & 3.775 & 2.917 & 15.761 & 1.248 & 0.0170 & 12 & 3.753 & 0.0446 \\ 
A87 & 0.1031 & 1.85 & 6734 & 3.775 & 2.917 & 15.761 & 1.248 & 0.0170 & 20 & 3.753 & 0.0888 \\ 
A88 & 0.1031 & 1.85 & 6734 & 3.775 & 2.917 & 15.761 & 1.248 & 0.0170 & 19 & 3.753 & 0.0952 \\ 
A89 & 0.1031 & 1.85 & 6734 & 3.775 & 2.917 & 15.761 & 1.248 & 0.0170 & 11 & 3.753 & 0.0497 \\ 
A90 & 0.1031 & 1.85 & 6734 & 3.775 & 2.917 & 15.761 & 1.248 & 0.0170 & 13 & 3.753 & 0.0484 \\ 
A91 & 0.1031 & 1.85 & 6734 & 3.775 & 2.917 & 15.761 & 1.248 & 0.0170 & 14 & 3.753 & 0.0610 \\ 
A92 & 0.1031 & 1.85 & 6734 & 3.775 & 2.917 & 15.761 & 1.248 & 0.0170 & 15 & 3.753 & 0.0788 \\ 
A93 & 0.1031 & 1.85 & 6734 & 3.775 & 2.917 & 15.761 & 1.248 & 0.0170 & 10 & 3.753 & 0.0635 \\ 
A94 & 0.1031 & 1.85 & 6734 & 3.775 & 2.917 & 15.761 & 1.248 & 0.0170 & 16 & 3.753 & 0.0816 \\ 
A95 & 0.1031 & 1.85 & 6734 & 3.775 & 2.917 & 15.761 & 1.248 & 0.0170 & 18 & 3.753 & 0.0903 \\ 
A96 & 0.1031 & 1.85 & 6734 & 3.775 & 2.917 & 15.761 & 1.248 & 0.0170 & 9 & 3.753 & 0.0736 \\ 
A97 & 0.1031 & 1.85 & 6734 & 3.775 & 2.917 & 15.761 & 1.248 & 0.0170 & 17 & 3.753 & 0.0851 \\ 
A98 & 0.1137 & 1.87 & 6743 & 3.777 & 2.926 & 15.947 & 1.220 & 0.0180 & 17 & 3.753 & 0.0379 \\ 
A99 & 0.1137 & 1.87 & 6743 & 3.777 & 2.926 & 15.947 & 1.220 & 0.0180 & 16 & 3.753 & 0.0432 \\ 
A100 & 0.1137 & 1.87 & 6743 & 3.777 & 2.926 & 15.947 & 1.220 & 0.0180 & 18 & 3.753 & 0.0363 \\ 
A101 & 0.1137 & 1.87 & 6743 & 3.777 & 2.926 & 15.947 & 1.220 & 0.0180 & 15 & 3.753 & 0.0390 \\ 
A102 & 0.1137 & 1.87 & 6743 & 3.777 & 2.926 & 15.947 & 1.220 & 0.0180 & 20 & 3.753 & 0.0318 \\ 
A103 & 0.1137 & 1.87 & 6743 & 3.777 & 2.926 & 15.947 & 1.220 & 0.0180 & 19 & 3.753 & 0.0454 \\ 
A104 & 0.1180 & 1.88 & 6781 & 3.780 & 2.925 & 16.301 & 1.197 & 0.0180 & 15 & 3.751 & 0.0972 \\ 
A105 & 0.1180 & 1.88 & 6781 & 3.780 & 2.925 & 16.301 & 1.197 & 0.0180 & 20 & 3.751 & 0.0636 \\ 
A106 & 0.1180 & 1.88 & 6781 & 3.780 & 2.925 & 16.301 & 1.197 & 0.0180 & 19 & 3.751 & 0.0643 \\ 
A107 & 0.1180 & 1.88 & 6781 & 3.780 & 2.925 & 16.301 & 1.197 & 0.0180 & 16 & 3.751 & 0.0735 \\ 
A108 & 0.1180 & 1.88 & 6781 & 3.780 & 2.925 & 16.301 & 1.197 & 0.0180 & 18 & 3.751 & 0.0662 \\ 
A109 & 0.1180 & 1.88 & 6781 & 3.780 & 2.925 & 16.301 & 1.197 & 0.0180 & 17 & 3.751 & 0.0620 \\ 
A110 & 0.1112 & 1.87 & 6807 & 3.780 & 2.917 & 16.465 & 1.200 & 0.0170 & 11 & 3.750 & 0.0629 \\ 
A111 & 0.1112 & 1.87 & 6807 & 3.780 & 2.917 & 16.465 & 1.200 & 0.0170 & 9 & 3.750 & 0.0594 \\ 
A112 & 0.1112 & 1.87 & 6807 & 3.780 & 2.917 & 16.465 & 1.200 & 0.0170 & 17 & 3.750 & 0.0579 \\ 
A113 & 0.1112 & 1.87 & 6807 & 3.780 & 2.917 & 16.465 & 1.200 & 0.0170 & 12 & 3.750 & 0.0647 \\ 
A114 & 0.1112 & 1.87 & 6807 & 3.780 & 2.917 & 16.465 & 1.200 & 0.0170 & 14 & 3.750 & 0.0716 \\ 
A115 & 0.1112 & 1.87 & 6807 & 3.780 & 2.917 & 16.465 & 1.200 & 0.0170 & 13 & 3.750 & 0.0556 \\ 
A116 & 0.1112 & 1.87 & 6807 & 3.780 & 2.917 & 16.465 & 1.200 & 0.0170 & 15 & 3.750 & 0.0720 \\ 
A117 & 0.1112 & 1.87 & 6807 & 3.780 & 2.917 & 16.465 & 1.200 & 0.0170 & 10 & 3.750 & 0.0752 \\ 
A118 & 0.1112 & 1.87 & 6807 & 3.780 & 2.917 & 16.465 & 1.200 & 0.0170 & 20 & 3.750 & 0.0752 \\ 
A119 & 0.1112 & 1.87 & 6807 & 3.780 & 2.917 & 16.465 & 1.200 & 0.0170 & 18 & 3.750 & 0.0594 \\ 
A120 & 0.1112 & 1.87 & 6807 & 3.780 & 2.917 & 16.465 & 1.200 & 0.0170 & 19 & 3.750 & 0.0661 \\ 
A121 & 0.1214 & 1.89 & 6817 & 3.782 & 2.926 & 16.660 & 1.174 & 0.0180 & 17 & 3.750 & 0.0981 \\ 
A122 & 0.1214 & 1.89 & 6817 & 3.782 & 2.926 & 16.660 & 1.174 & 0.0180 & 18 & 3.750 & 0.0946 \\ 
A123 & 0.1214 & 1.89 & 6817 & 3.782 & 2.926 & 16.660 & 1.174 & 0.0180 & 20 & 3.750 & 0.0980 \\ 
A124 & 0.1214 & 1.89 & 6817 & 3.782 & 2.926 & 16.660 & 1.174 & 0.0180 & 19 & 3.750 & 0.0860 \\ 
A125 & 0.1146 & 1.88 & 6843 & 3.782 & 2.919 & 16.833 & 1.179 & 0.0170 & 14 & 3.749 & 0.0517 \\ 
A126 & 0.1146 & 1.88 & 6843 & 3.782 & 2.919 & 16.833 & 1.179 & 0.0170 & 15 & 3.749 & 0.0508 \\ 
A127 & 0.1146 & 1.88 & 6843 & 3.782 & 2.919 & 16.833 & 1.179 & 0.0170 & 13 & 3.749 & 0.0709 \\ 
A128 & 0.1146 & 1.88 & 6843 & 3.782 & 2.919 & 16.833 & 1.179 & 0.0170 & 16 & 3.749 & 0.0683 \\

\end{longtable}


\bibliography{sample631}{}
\bibliographystyle{aasjournal}



\end{document}